\documentclass[12pt]{article}
\usepackage[english]{babel}
\usepackage{amsmath,amssymb}

\begin{document}

\newtheorem{theorem}{\bf Theorem}[section]
\newtheorem{lemma}[theorem]{\bf Lemma}
\newtheorem{corollary}[theorem]{\bf Corollary}
\newtheorem{proposition}[theorem]{\bf Proposition}
\newtheorem{definition}[theorem]{\bf Definition}
\newtheorem{conjecture}[theorem]{\bf Conjecture}
\newtheorem{assumption}[theorem]{\bf Assumption}
\def\remark{\noindent {\bf Remark.}\ }
\renewcommand{\theequation}{\arabic{section}.\arabic{equation}}

\def\Assum#1{\bf Assumption \rm (#1).\it}
\def\Defin#1{\bf Definition #1.\it}
\def\Lemma#1{\bf Lemma #1.\it}
\def\Propo#1{\bf Proposition #1.\it}
\def\Theor#1{\bf Theorem #1.\it}
\def\Conse#1{\bf Corollary #1.\it}
\def\proof{\noindent\bf Proof. \rm}
\def\remar{\bf Remark. \rm}
\def\thank{\bf Acknowledgment. \rm}

\def\hrm#1{\hbox{\rm #1}}
\def\hit#1{\hbox{\it #1}}
\def\hbf#1{\hbox{\bf #1}}
\def\vynos#1{\par\hangindent\parindent\indent\llap{#1\enspace}\ignorespaces}

\def\al{\alpha}          \def\be{\beta}
\def\ga{\gamma}          \def\de{\delta}
\def\ve{\varepsilon}     \def\ep{\epsilon}
\def\ze{\zeta}           \def\et{\eta}
\def\th{\theta}          \def\vt{\vartheta}
\def\io{\iota}           \def\im{\imath}
\def\jm{\jmath}          \def\ka{\kappa}
\def\la{\lambda}         \def\vpi{\varpi}
\def\rh{\rho}            \def\vr{\varrho}
\def\si{\sigma}          \def\vs{\varsigma}
\def\ta{\tau}            \def\up{\upsilon}
\def\ph{\phi}            \def\vp{\varphi}
\def\ch{\chi}
\def\ps{\psi}            \def\om{\omega}

\def\Ga{\Gamma}          \def\De{\Delta}
\def\Th{\Theta}          \def\La{\Lambda}
\def\Si{\Sigma}          \def\Up{\Upsilon}
\def\Ph{\Phi}            \def\Ps{\Psi}
\def\Om{\Omega}

\def\R{\mathbb{R}}          \def\C{\mathbb{C}}
\def\N{\mathbb{N}}          \def\Z{\mathbb{Z}}

\def\cE{{\cal E}}

\def\lar{\leftarrow}      \def\rar{\rightarrow}
\def\Lar{\Leftarrow}      \def\Rar{\Rightarrow}

\def\pr{\prime}          \def\pa{\partial}
\def\ti{\tilde}          \def\ol{\overline}
\def\fa{\forall}         \def\s-{\setminus}
\def\sub{\subset}        \def\sups{\supset}
\def\n={\not=}           \def\eq{\equiv}
\def\ex{\exists}         \def\ty{\infty}
\def\vn{\varnothing}

\def\rd{\buildrel\rm def \over =}

\def\varlimsup{\mathop{\overline{\hbox{\rm lim}}}}
\def\varliminf{\mathop{\underline{\hbox{\rm lim}}}}
\def\Re{\mathop{\rm Re}\nolimits}
\def\Im{\mathop{\rm Im}\nolimits}
\def\tr{\mathop{\rm tr}\nolimits}
\def\mes{\mathop{\rm mes}\nolimits}
\def\dom{\mathop{\rm Dom}\nolimits}
\def\im{\mathop{\rm Im}\nolimits}
\def\rang{\mathop{\rm rang}\nolimits}
\def\supp{\mathop{\rm supp}\nolimits}
\def\cot{\mathop{\rm ctg}\nolimits}
\def\diam{\mathop{\rm diam}\nolimits}
\def\min{\mathop{\rm min}\nolimits}
\def\dist{\mathop{\rm dist}\nolimits}
\def\cosh{\mathop{\rm ch}\nolimits}
\def\sinh{\mathop{\rm sh}\nolimits}
\def\tanh{\mathop{\rm th}\nolimits}
\def\coth{\mathop{\rm cth}\nolimits}
\def\Ln{\mathop{\rm Ln}\nolimits}
\def\diag{\mathop{\rm diag}\nolimits}
\def\const{\mathop{\rm Const}\nolimits}
\def\res{\mathop{\rm res}\nolimits}
\def\sign{\mathop{\rm sign}\nolimits}
\def\varlimsup{\mathop{\overline{\hbox{\rm lim}}}}
\def\varliminf{\mathop{\underline{\hbox{\rm lim}}}}

\def\vu{{\bf u}}
\def\vf{{\bf f}}
\def\val{\vec{\al}}

\title{Averaging in scattering problems}

\author{Buslaev Vladimir S.\footnote{St. Petersburg State University, Physical Faculty,
Department of Mathematical Physics, St. Petersburg, Russia; e-mail: buslaev@mph.phys.spbu.ru.}\ \
and Pozharskii Alexey A.\footnote{St. Petersburg State University, Physical Faculty,
Department of Mathematical Physics, St. Petersburg, Russia; e-mail: pozharsky@math.nw.ru.}
\thanks{The work was partially supported by the grants RFBR 08-01-00209-a and 07-01-92169.}}

\maketitle

\begin{abstract}
We consider the scattering that is described by the equation $(-\Delta_x + q(x,\frac{x}{\epsilon}) - E)\psi=
f(x), \psi = \psi(x,\epsilon) \in \C, x \in \R^d, \epsilon > 0, E > 0,$ where $q(x,y)$ is a periodic function of $y$, $q$ and $f$ have compact supports with respect to $x$. We are interested in the solution satisfying the radiation condition at infinity and describe the asymptotic behavior of the solution as $\epsilon \to 0$.
In addition, we find the asymptotic behavior of the scattering amplitude of the plain wave.
Either of them (the solution and the amplitude) in the leading orders are described by the averaged equation
with the potential $$\hat{q}(x) = \frac{1}{|\Omega|}\int_{\Omega}q(x,y)dy.$$
\end{abstract}


\section{Introduction}

Already a long period of time mathematicians are actively studying various and important for applications models of media which proporties change rapidly and periodically with space coordinates, see for example, [1-4,6-8]. As a rule (but not always!), such media respond to any static loading as media with some averaged constant characteristics. Physically, also another situation could be interesting. Media with rapidly changing properties can be local perturbation of homogeneous media and can be subjucted to external exposure. In this situation it is interesting to describe the characteristics of scattering that is generated by the mentioned local perturbation. Our goal at this work is to show that the characteristics of the scattering in certain conditions also can behave as the characteristics of some averaged media.

We consider the model described by the differential operator
\begin{equation}
\label{HomoMain}
H_\ve = - \Delta_x + q\left(x, \frac{x}{\ve}\right), \quad x  \in \R^d,
\end{equation}
where $\Delta_x$ is the Laplacian with respect to the variable  $x$ and $\ve$ is a small positive parameter.
As for the function $q = q(x, y)$, $x, y \in \R^d$ it is supposed that the following assumption is satisfied.

\begin{assumption}\label{Asum1}\
\begin{enumerate}
\item[i)]
$q$ is a real valued function of the class функция  $C^{\ty}(\R^{2d})$;

\item[ii)]
$q(x, y) = 0$ for $|x| \geqslant R$, $y\in\R^d$, where $R$ is some positive number;

\item[iii)]
$q(x, y)$ is periodic with respect to   $y$ (with some periodicity cell $\Om$).
\end{enumerate}
\end{assumption}

\medskip

For more detailed description of the setting of the problem and the main results we will have to recall certain standard for the situation definitions.

Let $H_\ga^s$, $s = 0, 1, 2, \ldots$, $\ga\in\R$, or rather more detailed $H_\ga^s(\R^d)$, be a Sobolev space of functions $\R^d \rar \C$ with the following norm
\begin{equation*}
{\| \ps \|}^2_{H_\ga^s} = \sum_{|k| \leqslant s}\
\int\limits_{\R^d} \left|
\frac{\pa^{|k|} \ps(x)}{\pa x^k}
\right|^2 \left(1 + |x|^2 \right)^{-\ga} \, d x.
\end{equation*}

Let $\ps$ be a solution of the equation
\begin{equation}\label{MainEqPF}
\left(- \Delta_x + p(x) - E \right) \ps(x) = f(x),\quad x \in \R^d,
\end{equation}
where $p$ and $f$ are functions with compact supports and  $E>0$.
Let us tell that $\ps$ satisfies \emph{radiation conditions} if
\begin{equation}\label{CondZomm}
\ps(x) = T(\hat{x}) \frac{e^{i\sqrt{E}\, |x|}}{|x|^{\frac{d-1}{2}}}
+ o\left( \frac{1}{|x|^{\frac{d-1}{2}}} \right), \quad
\hat{x} = \frac{x}{|x|}, \quad
|x| \rar \ty.
\end{equation}
Here $T$ is undefined at the moment smooth function  $S^{d-1} \rar \C$.

Let $\ps$ be a solution of equation  (\ref{MainEqPF}) where $p$ is an arbitrary function with compact support,
$f \in H_{-\ga}^s$ with some $\ga > 1$ and $E>0$.
Let us tell that $\ps$ satisfies the \emph{radiation conditions in a weak sense} if
$\ps \in H_{\ga}^{s+2}$ and exists a sequence of functions  $f_n \in H_{-\ga}^s$ with compact supports  converging to $f$ in the space $H_{-\ga}^s$ as $n \rar \ty$, such that
\begin{equation*}
\lim_{n \rar \ty} \| \ps_n - \ps \|_{H_{\ga}^{s + 2}} = 0,
\end{equation*}
where $\ps_n$ is the solution of the equation \ref{MainEqPF} with the right hand side $f_n$ satisfying the radiation condition.

Further we will need the following result, see \cite{Va,He}.

\begin{theorem}\label{LmEstHhat}
Let $p$ be a smooth function with a compact support, $E > 0$, $s\geqslant 0$, $\ga > 1$ and $f \in H_{-\ga}^{s}$.
Then there exists a unique solution of equation  \hrm{(\ref{MainEqPF})},
satisfying the radiation condition in a weak sense.
Such solution allows the estimate
\begin{equation}\label{EstPsGa}
\| \ps\|_{H_{\ga}^{s+2}} \leqslant C \| f \|_{H_{-\ga}^{s}},
\end{equation}
where the constant $C$ does not depend on  $f$.
\end{theorem}

Theorem
\ref{LmEstHhat}
implies that the operator $(-\De_x + p(x) - E)$ supplemented by the radiation condition in a weak sense, is formally invertible. Let us denote the inverse operator by
$(-\De_x + p(x) - E - i0)^{-1}$
(it is known that the solution satisfying the radiation condition  can be obtained from the the solution decreasing at infinity in the limit $\Im E \rar +0$,  $\Im E > 0$).
Estimate (\ref{EstPsGa}) guarantees that $(-\De_x + p(x) - E - i0)^{-1}$ acts as an bounded operator from
$H_{-\ga}^{s}$ to $H_{\ga}^{s+2}$, for $\ga > 1$ and $s \geqslant 0$.

We consider, first of all, the solution of the equation
\begin{equation}
\label{HomoEquat}
(H_\ve - E) \ps(x) = f(x),
\end{equation}
assuming that $E$ is a fixed positive parameter аnd $f$ is a given function of the class $C^{\ty}_0(\R^{d})$.
The solution $\ps$ is determined by the radiation condition  (\ref{CondZomm}).

The main results of the work are founded on the fact that problem  (\ref{HomoEquat}), (\ref{CondZomm}) allows the formal solution
\begin{equation*}
\ps(x, \ve) = \sum_{n \geqslant 0} \ve^n \ps_n \left(x, \frac{x}{\ve}\right),
\end{equation*}
where the coefficients  $\ps_n(x, y)$ depend on $y$ periodically and satisfy in an appropriate sense the radiation condition.  This solution is constructed in the work explicitely, see theorem~\ref{LmAsimpRaw}.

One of the main results of the work is the asymptotic description of the operator
$(H_\ve - E - i0)^{-1}$ as $\ve \rar +0$, see theorem~\ref{MainThR}.
Here we formulate the simplest consequence of theorem~\ref{MainThR} that describes the leading order of the asymptotic expansion of the solution.

\begin{theorem}\label{ThCol1}
Let the potential $q$ satisfy the assumption
\hrm{\ref{Asum1}},
 $E > 0$ and $\ga > 1$.
Then the estimate $\ve > 0$ holds
\begin{equation*}
\| (H_\ve - E - i0)^{-1} - (\hat{H} - E - i0)^{-1}
\|_{H_{-\ga}^{0} \rar H_{\ga}^{1}}
\leqslant C \ve.
\end{equation*}
Here $\hat{H}$ is the averaged operator
\begin{equation*}
\hat{H} = -\Delta_x + \hat{q}(x), \quad \hat{q}(x) = \frac{1}{|\Om|}\int\limits_\Om q(x, y) \, d y
\end{equation*}
and the constant $C$ does not depend on $\ve$ (but can depend on $E$ and $\ga$).
\end{theorem}

Theorem~\ref{MainThR}  almost immediately leads to a consequence  that can be considered as the main result of the work.

Let $F_\ve(\hat{x}, \ka)$, $k \in \R^d$, $|\ka| = E$ be the scattering amplitude of the plain
wave $e^{i <x,\ka>}$,
that is defined by the equation
\begin{equation*}
(H_\ve - E) \ps(x) = 0
\end{equation*}
and the asymptotic expansion at infinity
\begin{equation}\label{ReflCondForFlatWave}
\ps(x) = e^{i <x,\ka>} + F_\ve(\hat{x}, \ka) \frac{e^{i\sqrt{E}\, |x|}}{|x|^{\frac{d-1}{2}}}
+ o\left( \frac{1}{|x|^{\frac{d-1}{2}}} \right), \quad
|x| \rar \ty.
\end{equation}
We describe the asymptotic behavior of the amplitude  $F_\ve(\hat{x}, \ka)$ as $\ve \rar +0$, see theorem~\ref{MainTA}.
Here we formulate the consequence describing the leading order of the asymptotic expansion.

\begin{theorem}\label{ThCol2}
Let the potential $q$ satisfy assumption  \hrm{\ref{Asum1}} and $E > 0$.
Then the following estimate holds
\begin{equation*}
\sup_{\hat{x}, \ka}
\left| F_\ve(\hat{x}, \ka) - \hat{F}_0(\hat{x}, \ka) \right| \leqslant C \ve,
\end{equation*}
where $\hat{F}_0(\hat{x}, \ka)$ is the scattering amplitude for the averaged operator $\hat{H}$ and the constant $C$ does not depend on ~$\ve$.
\end{theorem}

\section{Formal asymptotic solutions}

This section is devoted to the constructing of formal asymptotic solutions, as $\ve\rar 0$,
of the equation
\begin{equation}
\label{HomoMainF}
- \Delta_x \ps + q\left(x, \frac{x}{\ve}\right) \ps - E \ps = f(x),
\quad x \in \R^d.
\end{equation}
We will suppose that $E > 0$, $f \in C_0^{\ty}(\R^{d})$ and that the solution  $\ps$ satisfies  radiation condition   (\ref{CondZomm}) at infinity.

To separate the slow and fast dependencies on the argument we will seek the solution of equation  (\ref{HomoMainF}) in the form
\begin{equation}
\label{MainRep}
\ps = \Ps\left(x, \frac{x}{\ve}, \ve\right).
\end{equation}
It is easy to check by the direct substitution that, if the function
функция $\Ps(x, y, \ve)$ satisfies the equation
\begin{equation}\label{MainAdia}
- \Delta_y \Ps - 2 \ve \nabla_y \nabla_x \Ps + \ve^2 (-\Delta_x + q(x, y) - E) \Ps = \ve^2 f(x).
\end{equation}
then the function $\ps = \Ps(x, x/\ve, \ve)$ satisfies equation  (\ref{HomoMainF}). Thus, the constucting of
formal solutions of equation  (\ref{HomoMainF}) is reduced to the constructing of formal solutions of equation  (\ref{MainAdia}).

The formal solutions of equation  (\ref{MainAdia}) will be seeked in the form
\begin{equation}
\label{ReprAsympU}
\Ps(x, y, \ve) = \sum_{n\geqslant 0} \ve^n \Ps_n(x, y),
\end{equation}
and it is supposed here that functions $\Ps_n(x, y)$ are $\Om$ periodic with respect to $y$.

As for  radiation conditions  (\ref{CondZomm}), they, naturally, have to be interpreted by the following way:  each coefficient $\Ps_n(x, y)$ for sufficiently large  $x$ does not  depend on $y$, and has to satisfy the radiation condition as $x \rar \ty$.

Plugging expansion  (\ref{ReprAsympU}) in equation (\ref{MainAdia}), and comparing the coefficients corresponding to the equal powers of  $\ve$, we obtain the following recurrence system:
\begin{equation}\label{RekPs}
\Delta_y \Ps_n =
- 2 \nabla_y \nabla_x \Ps_{n-1} +
(-\Delta_x + q(x, y) - E) \Ps_{n-2} - \de_{n2} f(x), \quad
n\geqslant 0,
\end{equation}
where $\Ps_{-2} \eq 0$ и $\Ps_{-1} \eq 0$.

It is reasonable to separate from the function $\Ps_n(x, y)$ orthogonal in  $L^2(\Om)$
(in the sense of dependence on $y$) components:
\begin{equation}
\label{ReprPs}
\Ps_n(x, y) = \vp_n(x, y) + \ps_n(x),
\end{equation}
where  $\vp(x, y)$ -- $\Om$ is a periodic function on  $y$ satisfying the condition
\begin{equation}
\label{CondPerVp}
\int\limits_\Om \vp_n(x, y) \, d y = 0.
\end{equation}

The following Lemma can be proved.

\begin{lemma}\label{LmAsimpRaw}
Let $E > 0$, $f \in C_0^\ty(\R^d)$ and $q$ satisfies assumption \hrm{\ref{Asum1}}.
Then there exists a formal solution of equation
\hrm{(\ref{HomoMainF})} of the form \hrm{(\ref{ReprAsympU})},
\hrm{(\ref{ReprPs})},  \hrm{(\ref{CondPerVp})}.
Functions
$\vp_n \in C^\ty(\R^{2d})$ и $\ps_n \in C^\ty(\R^d)$
can be found from the following recurrence relations
\begin{multline}\label{RekVpn}
\Delta_y \vp_n(x, y) =
- 2 \nabla_y \nabla_x \vp_{n-1}(x, y) + \\
+ (-\Delta_x + q(x, y) - E) \Ps_{n-2}(x, y) - \de_{n2} f(x), \quad
n\geqslant 0,
\end{multline}
\begin{equation}\label{RekPsn}
(\hat{H} - E) \ps_{n}(x) =
\de_{n0} f(x) -
\frac{1}{|\Om|} \int\limits_\Om q(x, y) \vp_{n}(x, y) \, d y, \quad
n\geqslant 0.
\end{equation}
Here
\begin{equation*}
\hat{H} = -\Delta_x + \hat{q}(x), \quad
\hat{q}(x) = \frac{1}{|\Om|}\int\limits_\Om q(x, y) \, d y,
\end{equation*}
$\ps_n$ satisfy radiation condition  \hrm{(\ref{CondZomm})} and
$\vp_n(x, y) \eq 0$ for $|x| \geqslant R$, $y\in\R^d$.
\end{lemma}

\proof
Plugging (\ref{ReprPs}) in (\ref{RekPs}), one obtains for  the function $\vp_n$  equation of the form  (\ref{RekVpn}).
Equation (\ref{RekVpn}) has a unique periodic solution satisfying condition
(\ref{CondPerVp}) if and only if right hand side  (\ref{RekVpn})
is orthogonal to the constant function with respect to the scalar product in   $L_2(\Om)$. It is easy to see that the orthogonality conditions can be written in the form
\begin{equation}\label{RekPsnNN}
(\hat{H} - E) \ps_{n-2}(x) =
\de_{n2} f(x) -
\frac{1}{|\Om|} \int\limits_\Om q(x, y) \vp_{n-2}(x, y) \, d y, \quad
n\geqslant 2.
\end{equation}
For $n = 0, 1$ equations (\ref{RekVpn}) have solutions unconditionally.

Now we have to consider recurrence system  (\ref{RekVpn}), (\ref{RekPsnNN})
and have to make sure that, indeed, the system has the unique solution with the claimed properties.
From (\ref{RekVpn}) it follows that $\vp_0(x, y) \eq 0$. Then (\ref{RekPsnNN}) shows that
\begin{equation}\label{RekPs00}
(\hat{H} - E) \ps_0(x) = f(x).
\end{equation}
Equation (\ref{RekPs00}) has a unique solution satisfying radiation condition  (\ref{CondZomm}).
From $f \in C^\ty(\R^d)$ and $q \in C_0^\ty(\R^{2d})$ it follows that $\ps_0 \in C^\ty(\R^d)$.

Let us discuss now the continuation of the recurrence procedure. Assume that there are defined the functions Предположим, что определены функции
$\vp_0, \ldots, \vp_l$ and $\ps_0, \ldots, \ps_l$ that possess the properties claimed above. Consider equation
(\ref{RekVpn}) for $n = l + 1$.
Orthogonality condition  (\ref{RekPsnNN}) is satisfied authomatically, therefore the equation have the solution for   $n = l + 1$.
Equation (\ref{RekVpn}) for $n = l + 1$ can be rewritten in the form
\begin{multline*}
\Delta_y \vp_{l+1}(x, y) =
- 2 \nabla_y \nabla_x \vp_{l}(x, y) +
(-\Delta_x + q(x, y) - E) \vp_{l-1}(x, y) + \\
+ ( q(x, y) - \hat{q}(x) ) \ps_{l-1}(x) -
\frac{1}{|\Om|} \int\limits_\Om q(x, y) \vp_{l-1}(x, y) \, d y.
\end{multline*}
Now it is easy to see that $\vp_{l+1} \in C^\ty(\R^{2d})$ and $\vp_n(x, y) \eq 0$ for $|x| \geqslant R$, $y\in\R^d$.

Consider, at last, equation (\ref{RekPsnNN}) for $\ps_{l+1}$. The right hand side is known and is equal to zero for  $|x| \geqslant R$. Therefore the equation uniqually defines   $\ps_{l+1}$ that satisfies the radiation condition.
From $f \in C^\ty(\R^d)$, $q \in C_0^\ty(\R^{2d})$ and $\vp_{l+1} \in C^\ty(\R^{2d})$
it follows that $\ps_{l+1} \in C^\ty(\R^d)$.
Therefore, the recurrence procedure is satisfied.~$\square$

\medskip

\begin{remark}
The assumption $f \in C^\ty(\R^d)$ is necessary to construct all terms of the formal solution (\ref{HomoMainF}).
If we want to describe only several first terms of the series  (\ref{HomoMainF}), it will be necessary to assume only existence of several first derivative of
 $f$.
the compactness of the support of  $f$ is not also crucial and can be replaced by a suitable condition of decreasing it at infinity.
\end{remark}

\medskip

Two following lemmas describe important estimates for the components of formal solution (\ref{ReprAsympU}).

\begin{lemma}\label{LmEstimatePsnAdd}
Let $E > 0$, $s\geqslant 0$, $\ga > 1$ and $f \in H_{-\ga}^{s}$.
Then
\begin{equation}\label{EqLmEst1}
\max_{0 \leqslant |k| \leqslant 2} \max_{y \in \Om}
\left\| \frac{\pa^{|k|}}{\pa y^k} \Ps_n\left(\cdot, y\right) \right\|_{H_{\ga}^{s+4-n}}
\leqslant C \| f \|_{H_{-\ga}^{s}}
\end{equation}
for $2 \leqslant n \leqslant s +4$. Besides, $\Ps_1 \eq 0$,
$\Ps_0\left(x, y\right) = \ps_0(x)$ и
\begin{equation}\label{EqLmEst2}
\left\| \ps_0 \right\|_{H_{\ga}^{s+2}}
\leqslant C \| f \|_{H_{-\ga}^{s}},
\end{equation}
where the constant  $C$ does not depend on $f$.
\end{lemma}

\proof
It is easy to see that  $\vp_0 \eq 0$, $\vp_1 \eq 0$.
If $n=0$, equation (\ref{RekPsn}) can be rewritten in the form
\begin{equation*}
(\hat{H} - E) \ps_0 = f.
\end{equation*}
From this equation and from theorem~\ref{LmEstHhat} it follows (\ref{EqLmEst2}).
For $n=1$ right hand side (\ref{RekPsn}) is trivial and, therefore, $\ps_1 \eq 0$ and $\Ps_1 \eq 0$.
For $n=2$ equation (\ref{RekVpn}) takes on form
\begin{equation}\label{RekVp2}
\Delta_y \vp_2(x, y) = (q(x, y) - \hat{q}(x)) \ps_0(x).
\end{equation}
Noticing  $q \in C^\ty(\R^{2d})$ and (\ref{EqLmEst2}), one can easily obtain  that
\begin{equation}\label{EstVp2}
\max_{0 \leqslant |k| \leqslant 2} \max_{y \in \Om}
\left\| \frac{\pa^{|k|}}{\pa y^k} \vp_2\left(\cdot, y\right) \right\|_{H_{\ga}^{s+2}}
\leqslant C \| f \|_{H_{-\ga}^{s}}.
\end{equation}
It is clear that the constant  $C$ in (\ref{EstVp2}) depends on $q$, but not on $f$.
Now estimate (\ref{EqLmEst1}) for $n=2$ follows from (\ref{EstVp2}), (\ref{RekPsn}) and theorem~\ref{LmEstHhat}.
Estimates (\ref{EqLmEst1}) for $n \geqslant 3$ can be deduced by the induction from  (\ref{RekVpn}) and (\ref{RekPsn}).~$\square$

\begin{lemma}\label{LmEstimatePsn}
Let $E > 0$, $p\geqslant 2$, $\ga > 1$ and $f \in H_{-\ga}^{p-2}$.
Then
\begin{equation*}
\left\| \Ps_n\left(x, \frac{x}{\ve}\right) \right\|_{H_{\ga}^{2}}
\leqslant C \ve^{-2} \| f \|_{H_{-\ga}^{p-2}}
\end{equation*}
for $2 \leqslant n \leqslant p$. In addition, $\Ps_1 \eq 0$ and
\begin{equation*}
\left\| \Ps_0\left(x, \frac{x}{\ve}\right) \right\|_{H_{\ga}^{2}}
\leqslant C \| f \|_{H_{-\ga}^{p-2}}.
\end{equation*}
\end{lemma}

\proof
It is sufficient to notice that $\Ps_0\left(x, \frac{x}{\ve}\right) = \ps_0(x)$ and to use
lemma \ref{LmEstimatePsnAdd}.~$\square$

\section{Asymptotic expansion of the resolvent}

Consider a partial sum of formal series (\ref{ReprAsympU})
\begin{equation*}
\Ps^{(p)}(x, \ve) = \sum_{n=0}^{p} \ve^n \Ps_n\left(x, \frac{x}{\ve}\right).
\end{equation*}
Compute its discrepancy
\begin{equation*}
(H_\ve - E) \Ps^{(p)}(x, \ve) = f(x) + \ve^{p-1} Q^{(p)}(x, \ve), \quad p\geqslant 2,
\end{equation*}
where
\begin{multline*}
Q^{(p)}(x, \ve) =
(-\De_x + q(x, y) - E)
\left[ \vp_{p-1}(x, y) + \ve \vp_{p}(x, y) \right] - \\
- 2 \nabla_y \nabla_x \vp_{p}(x, y) +
( q(x, y) - \hat{q}(x) ) [\ps_{p-1}(x) + \ve \ps_{p}(x)] - \\
- \frac{1}{|\Om|} \int\limits_\Om q(x, z)
\left[ \vp_{p-1}(x, z) + \ve \vp_{p}(x, z) \right] \, d z
\biggr|_{y = x/\ve}.
\end{multline*}

It is clear that $\Ps^{(p)}(x, \ve)$ and $Q^{(p)}(x, \ve)$ depend on $f$ linearly. It is convenient to introduce the operators $A_p$ and $B_p$
\begin{equation*}
A_p f(x) = \Ps^{(p)}(x, \ve), \quad
B_p f(x) = Q^{(p)}(x, \ve), \quad p\geqslant 2.
\end{equation*}
From that it follows
\begin{equation}
\label{ResEqP}
(H_\ve - E) A_p = I + \ve^{p-1} B_p, \quad p \geqslant 2.
\end{equation}

\begin{lemma}\label{LmBoundApBp}
Let $E > 0$, $s\geqslant 0$ and $\ga > 1$.
Then for $p\geqslant 2$ the following estimates
\begin{equation}\label{EstAp}
\| A_p\|_{H_{-\ga}^{p-2} \rar H_{\ga}^{2}}
\leqslant C,
\end{equation}
\begin{equation}\label{EstBp}
\| B_p\|_{H_{-\ga}^{p-2} \rar H_{-\ga}^{0}}
\leqslant C,
\end{equation}
hold where the constant
$C$ can depend on $E$, $s$ and $\ga$, but not on $\ve$.
\end{lemma}

\proof
Estimate (\ref{EstAp}) follows from lemma~\ref{LmEstimatePsn}.
Lemma \ref{LmEstimatePsnAdd} implyes that
\begin{equation}\label{EstBpMin}
\| B_p\|_{H_{-\ga}^{p-2} \rar H_{\ga}^{0}}
\leqslant C_1,
\end{equation}
where  $C_1$ is some positive constant.
In its turn lemma
\ref{LmAsimpRaw} guarantees that the support of the function  $B_p f(x)$ for $f \in H_{-\ga}^{p-2}$
belongs to the ball $|x| < R$ .
Therefore there exists a constant $C_2$ such that the following estimate
\begin{equation}\label{EstBpMinPlus}
\| B_p f\|_{H_{-\ga}^{0}} \leqslant
C_2 \| B_p f\|_{H_{\ga}^{0}}
\end{equation}
holds.

Now (\ref{EstBp}) follows from  (\ref{EstBpMin}) and
(\ref{EstBpMinPlus}).~$\square$

How it was already mentioned in the Introduction, the operator  $H_\ve - E$ equipped by radiation condition  (\ref{CondZomm}) in weak sense for any $\ve > 0$ has a bounded inverse in a suitable space.
Let us denote the inverse operator by  $(H_\ve - E - i0)^{-1}$.

\begin{lemma}\label{LmBoundRes}
Let $E > 0$ and $\ga > 1$.
Then for sufficiently small  $\ve > 0$ there is true the estimate
\begin{equation*}
\| (H_\ve - E - i0)^{-1} \|_{H_{-\ga}^{0} \rar H_{\ga}^{2}} \leqslant C,
\end{equation*}
here the constant $C$ does not depend on $\ve$.
\end{lemma}

\proof
Let us use formula (\ref{ResEqP}) for $p=2$
\begin{equation}\label{EqForHveA2}
(H_\ve - E) A_2 = I + \ve B_2.
\end{equation}
Notice now that for sufficiently small $\ve$, as a consequence of  (\ref{EstBp}),
there is correctly defined the bounded operator
$(I + \ve B_2)^{-1}$ that acts in  $H_{-\ga}^{0}$. It allows the estimate
\begin{equation*}
\| (I + \ve B_2)^{-1} \|_{H_{-\ga}^{0} \rar H_{-\ga}^{0}}
\leqslant
\left(
1 - \ve \| B_2 \|_{H_{-\ga}^{0} \rar H_{-\ga}^{0}}
\right)^{-1} \leqslant C_1
\end{equation*}
with a constant  $C_1$ that does not depend on  $\ve$.
Now from  (\ref{EqForHveA2}) and (\ref{EstAp}) it follows that
\begin{equation}\label{RepRez2}
(H_\ve - E - i0)^{-1} =  A_2 (I + \ve B_2)^{-1},
\end{equation}
\begin{equation*}
\| (H_\ve - E - i0)^{-1} \|_{H_{-\ga}^{0} \rar H_{\ga}^{2}}
\leqslant
\| A_2 \|_{H_{-\ga}^{0} \rar H_{\ga}^{2}}
\| (I + \ve B_2)^{-1} \|_{H_{-\ga}^{0} \rar H_{-\ga}^{0}}
\leqslant C \not= C(\ve).
\ \square
\end{equation*}

\begin{theorem}\label{MainThR}
Let $E > 0$, $\ga > 1$ and $p\geqslant 2$.
Then for suffiently  $\ve>0$ the estimate
\begin{equation*}
\| (H_\ve - E - i0)^{-1} - A_{p}\|_{H_{-\ga}^{p-2} \rar H_{\ga}^{2}}
\leqslant C \ve^{p-1},
\end{equation*}
holds with a constant  $C$ that does not depend on  $\ve$.
\end{theorem}

\proof
From  (\ref{ResEqP}) it follows
\begin{equation*}
(H_\ve - E - i0)^{-1} - A_{p} =
 - \ve^{p-1} (H_\ve - E - i0)^{-1} B_{p},
\end{equation*}
where $p\geqslant 2$.
Lemmas \ref{LmBoundApBp} and \ref{LmBoundRes} implies that
\begin{equation*}
\| (H_\ve - E - i0)^{-1} B_{p} \|_{H_{-\ga}^{p-2} \rar H_{\ga}^{2}}
\leqslant
\| (H_\ve - E - i0)^{-1} \|_{H_{-\ga}^{0} \rar H_{\ga}^{2}}\
\| B_{p} \|_{H_{-\ga}^{p-2} \rar H_{-\ga}^{0}}
\leqslant C,
\end{equation*}
where $C$ does not depend on  $\ve$.
The proof is completed.~$\square$

\medskip

\noindent
{\bf Proof of theorem \ref{ThCol1}.}
For the proof it is sufficient to refer to theorem~\ref{MainThR}
for $p=2$, and notice that
$\| A_2 - A_0\|_{H_{-\ga}^{0} \rar H_{\ga}^{1}} \leqslant
C \ve$.~$\square$

\section{Asymptotic behavior of the scattering amplitude.}

Consider the homogeneous equation
\begin{equation}
\label{HomoMainO}
- \Delta_x \ps + q\left(x, \frac{x}{\ve}\right) \ps - E \ps = 0,
\quad x \in \R^d.
\end{equation}
As before we assume that  $E > 0$ and $q$ satisfies assumption~\ref{Asum1}.
There exists the solution $\ch(x, \ka)$ of equation (\ref{HomoMainO}) satisfying at infinity radiation condition ~(\ref{ReflCondForFlatWave}).
The solution $\ch$ describes the scattering of the plane wave  $e^{i<x,\ka>}$ by the potential  $q$. In this section we will be interested in the asymptotic behavior of scattering amplitude ~$F_\ve(\hat{x}, \ka)$ as $\ve \to 0$.

Following to already developed patterns, we seek the solution of equation  (\ref{HomoMainO}) in the form
\begin{equation}
\label{ReprAsympR}
\ch_p = \sum_{n=0}^p \ve^n \tilde{\Ps}_n\left(x, \frac{x}{\ve}\right),\quad
\tilde{\Ps}_n(t, y) = \tilde{\vp}_n(t, y) + \tilde{\ps}_n(t), \quad
\int\limits_\Om \tilde{\vp}_n(t, y) \, d y = 0.
\end{equation}

\begin{lemma}\label{LmAsimpRawZero}
Let  $E > 0$, $\ka \in\R^d$, $|\ka| = E$ and $q$ satisfies assumption  \hrm{\ref{Asum1}}.
Then there exist the formal solution of equation
\hrm{(\ref{HomoMainO})} of form \hrm{(\ref{ReprAsympR})}.
Then the solutions
$\tilde{\vp}_n \in C^\ty(\R^{2d})$ and $\tilde{\ps}_n \in C^\ty(\R^d)$
can be found from the following recurrence relations
\begin{equation}\label{RekVpnZ}
\Delta_y \tilde{\vp}_n(x, y) =
- 2 \nabla_y \nabla_x \tilde{\vp}_{n-1}(x, y) +
(-\Delta_x + q(x, y) - E) \tilde{\Ps}_{n-2}(x, y), \quad
n\geqslant 0,
\end{equation}
\begin{equation}\label{RekPsnZ}
(\hat{H} - E) \tilde{\ps}_{n}(x) =
- \frac{1}{|\Om|} \int\limits_\Om q(x, y) \tilde{\vp}_{n}(x, y) \, d y, \quad
n\geqslant 0.
\end{equation}
Here
\begin{equation*}
\hat{H} = -\Delta_x + \hat{q}(x), \quad
\hat{q}(x) = \frac{1}{|\Om|}\int\limits_\Om q(x, y) \, d y,
\end{equation*}
$\tilde{\ps}_0$ satisfies the radiation condition
\begin{equation*}
\left|
\tilde{\ps}_0(x, \ka) - e^{i<x,\ka>} - \hat{F}_0(\hat{x}, \ka) |x|^{\frac{1-d}{2}} e^{i |\ka| |x|}
\right| \leqslant C |x|^{\frac{-1-d}{2}},
\end{equation*}
$\tilde{\ps}_n$ satisfies the radiation condition
\begin{equation*}
\left|
\tilde{\ps}_n(x, \ka) - \hat{F}_n(\hat{x}, \ka) |x|^{\frac{1-d}{2}} e^{i |\ka| |x|}
\right| \leqslant C |x|^{\frac{-1-d}{2}},
\end{equation*}
for $n\geqslant 1$,
where $C$ is some constant that does not depend on $x$, and $\tilde{\vp}_n(x, y) \eq 0$ for
$|x| \geqslant R$, $y\in\R^d$.
\end{lemma}

\proof
We can give the proof that is analogous to the proof of lemma~\ref{LmAsimpRaw}.~$\square$

\begin{lemma}\label{LmDiffChp}
Let $E > 0$ and $q$ satisfy assumption \hrm{\ref{Asum1}}.
Then for $p \geqslant 0$ and $|x| \geqslant 2 T_0$
\begin{equation}\label{EstDifChp}
|\ch(x, \ka) - \ch_p(x, \ka)| \leqslant
C |x|^{\frac{1-d}{2}} \ve^{p+1},
\end{equation}
where  $C$ does not depend on  $\ve$ and $\ka$.
\end{lemma}

\proof
If $p \geqslant 2$ the direct computation shows that
\begin{equation*}
(H_\ve - E) \ch_p = \ve^{p-1} \Phi_p,
\end{equation*}
where
\begin{equation*}
\Phi_p(x, \ve) =
\De_y \left[ \tilde{\Ps}_{p+1}(x, y) + \ve \tilde{\Ps}_{p+2}(x, y) \right] +
2 \ve \nabla_x \nabla_y \tilde{\Ps}_{p+1}(x, y)
\biggr|_{y = x/\ve}.
\end{equation*}
It is easy to see that the function  $\Phi_p$ has a compact support that belongs to the ball $|x| < R$, and $\Phi_p\in C^{\ty}(\R^d)$.
Using a result that is analogous to the result of lemma~\ref{LmEstimatePsnAdd}
one can show that
\begin{equation}\label{EstPhp}
\| \Phi_p \|_{H_{-\ga}^{0}} \leqslant C \not= C(\ve).
\end{equation}
It is obvious that the function  $\ch_p - \ch$ satisfies the equation
\begin{equation*}
(H_\ve - E) (\ch_p - \ch) = \ve^{p-1} \Phi_p
\end{equation*}
and radiation condition  (\ref{CondZomm}).
Now referring to representation  (\ref{RepRez2}) one can obtain
\begin{equation}\label{ReChpCh}
\ch_p - \ch = \ve^{p-1} (H_\ve - E - i0)^{-1} \Phi_p =
\ve^{p-1} A_2 (I + \ve B_2)^{-1} \Phi_p.
\end{equation}
Let us consider the asymptotic behavior of the expression that enters to the right hand side of  (\ref{ReChpCh}),
for $|x| \rar \ty$.
from the definition of $B_2$ it is easy follows that the support of the function
$(I + \ve B_2)^{-1} \Phi_p$
belongs to the ball $|x| < T_0$.
Lemma \ref{LmBoundApBp} and estimate  (\ref{EstPhp}) imply that
\begin{equation}
\label{EstRS}
\| (I + \ve B_2)^{-1} \Phi_p \|_{H_{-\ga}^{0}}
\leqslant C \not= C(\ve).
\end{equation}
Notice now that the non-trivial contribution to the behavior of the function
$A_2 (I + \ve B_2)^{-1} \Phi_p$ for $|x| \rar \ty$
is described by the terms  $\ps_0(x)$ and $\ps_2(x)$
of the expansion of  $A_2$.
These terms are the solutions of the equation
\begin{equation*}
(\hat{H} - E) \ps_n = D_n,
\end{equation*}
where
\begin{equation*}
D_n = \de_{n0} (I + \ve B_2)^{-1} \Phi_p - \frac{1}{|\Om|} \int\limits_\Om q(x, y) \vp_{n}(x, y) \, d y,
\quad n = 0, 2.
\end{equation*}
It is easy to see that the support of  $F_n$ belongs to the ball  $|x| < R$.
From here and from  (\ref{EstRS}) it follows that  $D_n$ allows the following estimate
\begin{equation}
\label{EstF}
\| D_n \|_{H_{-\ga}^{0}} \leqslant C \not= C(\ve).
\end{equation}
As a result
\begin{equation}
\label{EstP}
\| \ps_n \|_{H_{\ga}^{0}} \leqslant C \not= C(\ve).
\end{equation}

It is easy to see that Легко видеть, что $\ps_0$ and $\ps_2$ satisfy to the integral equation
\begin{equation*}
\ps_n(x) =  \int\limits_{|y| \leqslant T_0} G_0(x, y; E)
\left(\rule{0ex}{2ex} \hat{q}(y) \ps_n(y) - D_n(y) \right) \, d y,
\end{equation*}
where $G_0(x, y; E)$ is the kernel of the resolvent $(-\De  - E - i0)^{-1}$
\begin{equation*}
G_0(x, y; E) =
\frac{i}{4} \left( \frac{\sqrt{E}}{2 \pi |x - y|}\right)^{\frac{d}{2} - 1}
H^{(1)}_{\frac{d}{2} - 1} \left(\sqrt{E}\, |x - y| \right)
\end{equation*}
and $H^{(1)}_{\frac{d}{2} - 1}$ is the Hankel function.
This implies that
\begin{equation*}
|\ps_n(x)| \leqslant
C\ |x|^{\frac{1-d}{2}}
\left(\max_{|y| \leqslant R} |\hat{q}(y)|
\int\limits_{|y| \leqslant R} |\ps_n(y)| \, d y
+
\int\limits_{|y| \leqslant R} |D_n(y)| \, d y
\right)
\end{equation*}
for $|x| \geqslant 2 R$, $n = 0, 2$.
Now taking into account  (\ref{EstF}) and (\ref{EstP}) we obtain
\begin{equation}\label{EstInftyPs}
|\ps_n(x)| \leqslant C |x|^{\frac{1-d}{2}},
\end{equation}
where $|x| \geqslant 2 R$ and  $C$ does not depend on $\ve$.
Now it is clear that for
$p\geqslant 2$ and $|x| \geqslant 2 R$ there is satisfied the following estimate
\begin{equation}\label{EstDiffCh}
|\ch(x, \ka) - \ch_p(x, \ka)| \leqslant C |x|^{\frac{1-d}{2}} \ve^{p-1}.
\end{equation}
Notice now that the terms  $\tilde{\Ps}_{p-1}$ and  $\tilde{\Ps}_p$ of expansion  (\ref{ReprAsympR})
allow the estimate
\begin{equation}\label{EstPsnPsn1}
\left|\tilde{\Ps}_{p-1}\left(x, \frac{x}{\ve}\right)\right| \leqslant
C |x|^{\frac{1-d}{2}} , \quad
\left|\tilde{\Ps}_{p}\left(x, \frac{x}{\ve}\right)\right| \leqslant
C |x|^{\frac{1-d}{2}},
\end{equation}
here $|x| \geqslant 2 R$ and $C$ does not depend on  $\ve$, $\ka$ and $x$.
These estimates can be obtained by the same constructions as (\ref{EstInftyPs}).

Now using (\ref{EstDiffCh}) and (\ref{EstPsnPsn1}) it is easy to show
\begin{equation*}\label{EstDiffCh-2}
|\ch(x, \ka) - \ch_{p-2}(x, \ka)| \leqslant C |x|^{\frac{1-d}{2}} \ve^{p-1},
\end{equation*}
for $p \geqslant 2$.
This concludes the proof.~$\square$

\begin{theorem}\label{MainTA}
Let $E > 0$ and $q$ satisfy  assumption \hrm{\ref{Asum1}}.
Then for $p \geqslant 0$ the following estimate
\begin{equation*}
\sup_{\hat{x}, \ka}
\left|
F_\ve(\hat{x}, \ka) - \sum_{n=0}^p \ve^n \hat{F}_n(\hat{x}, \ka)
\right| \leqslant C \ve^{p+1},
\end{equation*}
holds where $C$ does not depend on $\ve$.
\end{theorem}

\proof
From lemmas~\ref{LmAsimpRawZero} and~\ref{LmDiffChp} it follows that
\begin{equation}\label{EstDifChpLast}
\left|
F_\ve(\hat{x}, \ka) - \sum_{n=0}^p \ve^n \hat{F}_n(\hat{x}, \ka)
\right| \leqslant
C \ve^{p+1} + C_1(\ka, \ve) |x|^{-1},
\end{equation}
where the constant $C$ does not depend on  $x$, $\ka$ and $\ve$, and the constant $C_1(\ka, \ve)$ does not depend on $x$, but, may be, depends on $\ka$ and $\ve$.
To complete the proof it is sufficient to consider the limit  $|x| \rar \ty$ in estimate  (\ref{EstDifChpLast}).~$\square$

\medskip

\noindent
{\bf Proof of theorem \ref{ThCol2}}
follows from theorem  \ref{MainTA} for $p = 0$.~$\square$

\end{document}